\documentclass[%
 reprint,
 twocolumn,
 amsmath,amssymb,
 aps,
]{revtex4}

\usepackage{graphicx}
\usepackage{dcolumn}
\usepackage{bm}
\usepackage{hyperref}

\usepackage{longtable}
\begin{document}


\title{Activated and quantum creep of the charge-density waves in magnetic field in {\it o}-TaS$_3$}

\author{I.A. Cohn$^{1,2}$}
\author{S.V. Zaitsev-Zotov$^{1,2}$} 
\email{serzz@cplire.ru}
 \affiliation{$^1$Kotelnikov Institute of Radioengineering and Electronics of RAS, Mokhovaya 11, bld. 7, 125009 Moscow, Russia }
 \affiliation{$^{2}$HSE University, Faculty of Physics, 21/4 Staraya Basmannaya, 105066 Moscow, Russia}
             
\begin{abstract}

We demonstrate that magnetoresistance of creeping charge-density waves in the quasi-one dimensional conductor {\it o}-TaS$_3$ changes its character from a negative parabolic at $T\gtrsim 10$ K where it obeys $1/T^2$ law to a weakly temperature dependent negative nearly linear one at lower temperatures. The dominant contribution into the negative parabolic magnetoresistance comes from magnetic field induced  splitting of the CDW order parameter. The linear magnetoresistance arises due to CDW quantum interference similar to the scenario of negative linear magnetoresistance in single-electron systems. 
\end{abstract}

\maketitle


\section{Introduction}
Quantum tunneling, which consists in the penetration of the wave function through a potential barrier, is a remarkable manifestation of quantum mechanics. Macroscopic quantum tunneling (MQT) is understood as the tunneling penetration of the wave function corresponding to a large ensemble of particles. The probability of this penetration is reduced dramatically with the increase of the mass of penetrating system. As a result, although the tunneling of single electrons is well known and used in a variety of devices, the tunneling of a large number of particles at once has been discovered and studied only in a relatively small number of physical systems (as a review see \cite{takagi2002}). At the present, macroscopic quantum tunneling  is observed in small Josephson junctions \cite{MQT_sc}, thin superconductor wires \cite{Giordano},  vortices in superconductors \cite{Blatter}, magnetic systems \cite{MQT_spin}, spin- \cite{MQT_sdw} and charge-density waves (CDW)\cite{ZZqq}.

Peierls conductors are one of the few physical systems demonstrating collective conductivity due to the motion of a condensate - in this particular case, the CDW (as a recent review see \cite{review_Monceau}). 
The most studied materials in which this collective conductivity is realizes are inorganic quasi-one-dimensional conductors such as trichaclcogenides of transition metals MX$_3$ (M = Ta, Nb; X = Se, S, Te), blue bronzes M$_{0.3}$MoO$_3$ (M = K, Rb), (MX$_4$)$_n$I and NbS$_3$ (II). These materials are characterized by a relatively high value of the Peierls transition temperature $T_P\sim 10^2$ K. The values of the threshold field $E_T$ of the onset of nonlinear conductivity depend on type of material, its purity, and vary with temperature in the typical range $10^{-2}$-$10^3$~V/cm. Motion of the CDW is accompanied by generation of wide-band and narrow-band noises. The rich physics of inorganic CDW conductors is well documented and described in numerous reviews \cite{review_Monceau,Monceau_book,GrunerReview,ZZ_review}. 

The formation of the CDW below the Peierls transition temperature is accompanied by a decrease in the spin magnetic susceptibility, which is associated with a decrease in the concentration of single-particle excitations due to the formation of the Peierls gap \cite{sucseptibility}. This means that CDW formation is accompanied by singlet pairing of electrons. For this reason, the application of a magnetic field suppresses the Peierls transition \cite{suppression,Bjelis1999}. Analysis of the magnetic field effect on the spin degree of freedom with allowance for the spin-orbit interaction showed that the Zeeman splitting leads to a decrease in the temperature of the Peierls transition $\Delta T_P/T_P =-\gamma (\mu_BB/kT_P)^2$, where the parameter $\gamma \sim 1$ depends on the angle between the direction of the chains and the magnetic field \cite{suppression,Bjelis1999} and reaches its maximum value in a magnetic field parallel to the direction of the chains. This effect leads also to the appearance of the negative quadratic magnetoresistance 
\begin{equation}
\Delta \rho /\rho =-\frac{1}{2}\left(\frac{g\mu_BB}{kT}\right)^2
\label{eq:Ohmic}
\end{equation}
 in the Ohmic conduction regime \cite{tiedje}, where $g$ is $g$-factor omitted in Ref.~ \cite{tiedje}. While  the expected value of the effect in materials with $T_P = 100-300$~K in moderate magnetic fields $B\sim 5$~T does not exceed fractions of a percent, complete suppression of the Peierls transition is possible in materials with a much lower value of $T_P$ \cite{Graf}.
 
 Orthorhombic TaS$_3$ ({\it o}-TaS$_3$) is a quasi-one-dimensional conductor with the Peierls  transition temperature $T_P\approx 220$ K \cite{review_Monceau}. The energy gap opens at $T<T_P$ and covers the entire Fermi surface. 
The conductivity activation energy is 800-850 K in the temperature range $T_P /2 \leq T \leq T_P$ and is approximately halved at lower temperatures \cite{solitons}. This behavior is often associated with opening of an additional conduction channel provided by solitons \cite{solitons}. 

In {\it o}-TaS$_3$, low-temperature freezing-out of current carriers makes it possible to observe a collective contribution to the total conduction caused by slow motion of CDWs, known as CDW creep \cite{ZZqq}. In this regime, the CDW velocity is determined not by energy dissipation processes, as in sliding, but by the probability of overcoming of pinning barriers, which height depends on the electric field. The {\it I-V} curve is predicted to follow 
\begin{equation}
j = j_0\exp{ \left[ -\frac{T^*}{T}\left(\frac{ E_0}{E}\right) ^\mu \right] }
\label{eq:qq}
\end{equation}
law, where $\mu=1/2$ for three-dimensional pinning \cite{Nattermann}. This equation predicts nonlinear activated conduction with electric-field dependent  activation energy. 

In thin {\it o}-TaS$_3$ samples, when temperature decreases below 10~K, electric field-dependent activation energies goes to zero that indicates classical-to-quantum crossover in CDW creep, i.e. transition into the regime of temperature-independent penetration through pinning barriers \cite{ZZqq}. 
The current-voltage characteristic in this quantum creep regime follows $I \propto \exp [-(E_0/E )^{2} ]$ dependence.

This quantum creep observed initially in thin samples also exists in bulk pure samples of  {\it o}-TaS$_3$ where temperature-dependent nonlinear current approaches its final electric-field dependent value at $T\rightarrow 0$, whereas nonlinear conduction of impure samples goes to zero \cite{zeroTlimit}. The proposed creep mechanism implies quantum nucleation of CDW dislocation loops or dislocation vortex pair \cite{Duan,Maki,Hatakenaka,Matsukawa}. Further experimental and theoretical research has confirmed predictions on the shape of current-voltage characteristics which follows 
\begin{equation}
j=j_0(T) \exp{\left[-\left(\frac{E_0}{E}\right)^{\mu}\right]}.
\label{eq:qqimp}
\end{equation}
law with $\mu =1,2$ depending on pinning dimensionality \cite{Duan,Maki,Hatakenaka,Matsukawa,Inagaki}. More recent analysis of dynamics of an elastic medium driven through a random medium by a small applied force \cite{Gorokhov,Nattermann2} provides more reach range of possibilities including  Eq.~\ref{eq:qqimp} law with $\mu = 1/2$ for quantum creep of the CDW, in agreement with experimental observations in impure crystals \cite{zeroTlimit}. 

Here we present the experimental result of magnetoresistance study in pure and impure crystals {\it o}-TaS$_3$ in the helium temperature range in CDW creep regime. Our results reveal crossover between high-temperature negative parabolic magnetoresistance obeying $1/T^2$ law and negative linear  magnetoresistance which we attribute to quantum tunneling of the CDW through magnetic filed dependent pinning barriers.

\section{Experimental}

The crystals of {\it o}-TaS$_3$ have needle-like shape with very high aspect ratio up to few hundreds and even higher. The absence of the Shottky barrier between indium and {\it o}-TaS$_3$  together with very high aspect ratio allows to use 2-probe technique as a reliable method of transport measurements  for freshly prepared samples \cite{contacts}. Contacts to the samples were made by cold soldering with indium, which  provides stable low-resistance contacts and allows measurements in two-contact geometry in {\it o}-TaS$_3$. The results presented in this work were obtained in both nominally pure samples ($E_T(100 K)\lesssim 1$~V/cm) with typical cross-section area $s \sim 10{\rm - }10^3$~$\mu$m$^2$ and lengths $l\sim 0.2$ -- 5~mm and impure ones with $E_T(100 {\rm ~K})> 10^2$~V/cm and similar sizes. All the measurements were carried out with current flowing along the chains direction.

The measured relative variations of sample resistance in magnetic field are small and require  high stability of the sample temperature. We use Cernox$^\circledR$ temperature sensors together with temperature controllers providing a few mK temperature stability.  No temperature correction of the data due to magnetoresistance of  Cernox$^\circledR$ temperature sensor was taken into account. As typical values of temperature correction at 7 T do not exceeds several tens of mK \cite{cernox1,cernox2}, the respective accuracy of magnetoresistance measurements was limited to tenth of a percent at $T\leq 5$~K. 
The accuracy of current measurements were limited by  the value $10^{-13}$~A caused by vibrations generated by the closed cycle refrigerator.

\section{Results}
\subsection{Pure samples}
\begin{figure}
\includegraphics[width=8cm]{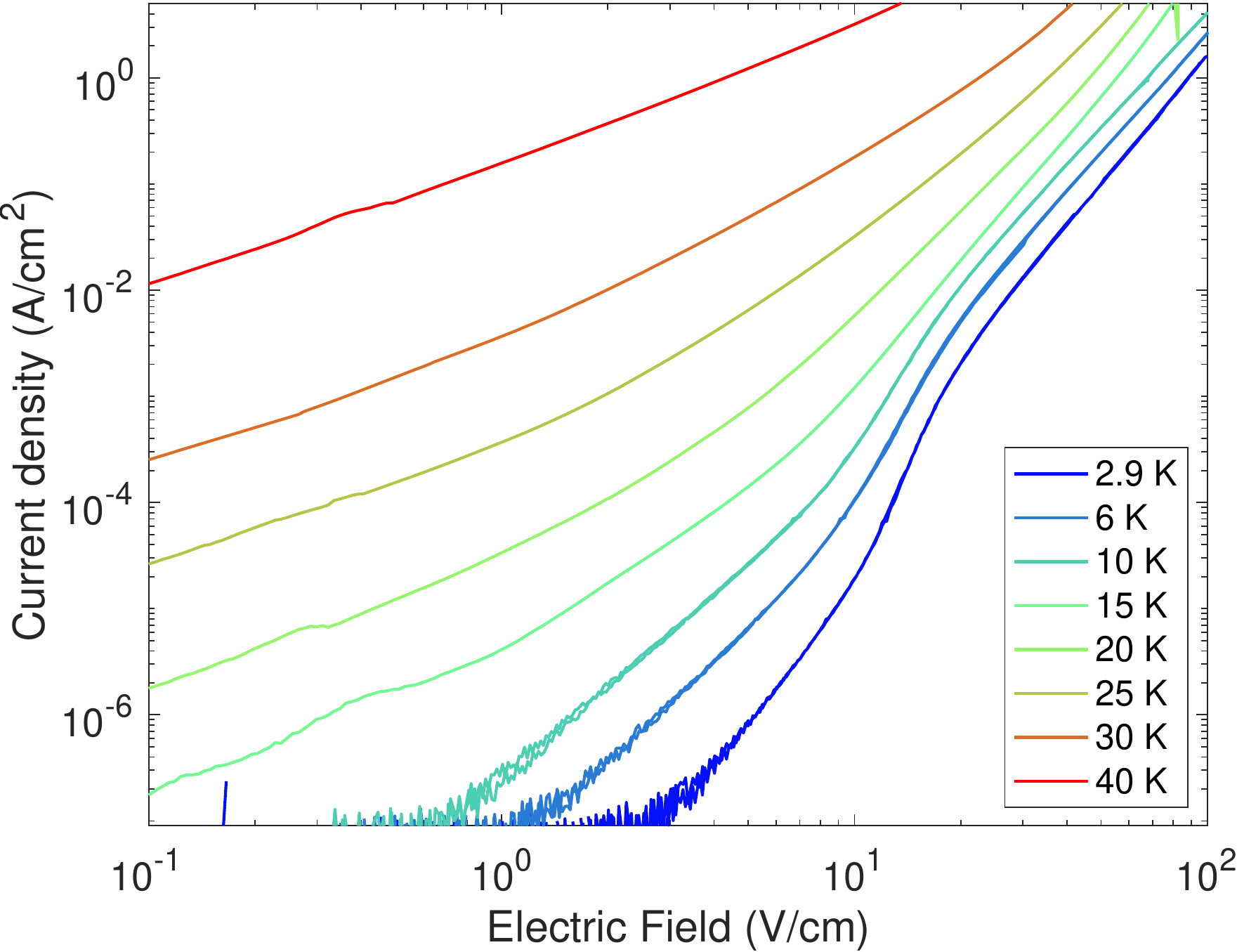} 
\caption{\label{fig:ivspure}Low-temperature set of {\it I-V} curves of the pure crystal of  {\it o}-TaS$_3$ sample in the CDW creep region.}
\end{figure}

CDW creep dominates in nonlinear conduction of {\it o}-TaS$_3$ at temperatures below 40-60 K \cite{ZZqq}. Fig.~\ref{fig:ivspure} shows a low-temperature set of current-voltage characteristics ({\it I-V} curves) of the pure sample with $E_T(120{\rm~K})=0.4$~V/cm. The curves are typical for {\it o}-TaS$_3$ and consist of sections of linear conductivity (flatter parts of the curves) and CDW creep region. The creep mode in pure crystals is characterized by high current noise corresponding to jumps of the CDW providing CDW progress in the direction of the electric field \cite{ZZqq}. This noise restricts the measurement accuracy of magnetoresistance especially in the low current region. 

\begin{figure}
\includegraphics[width=8cm]{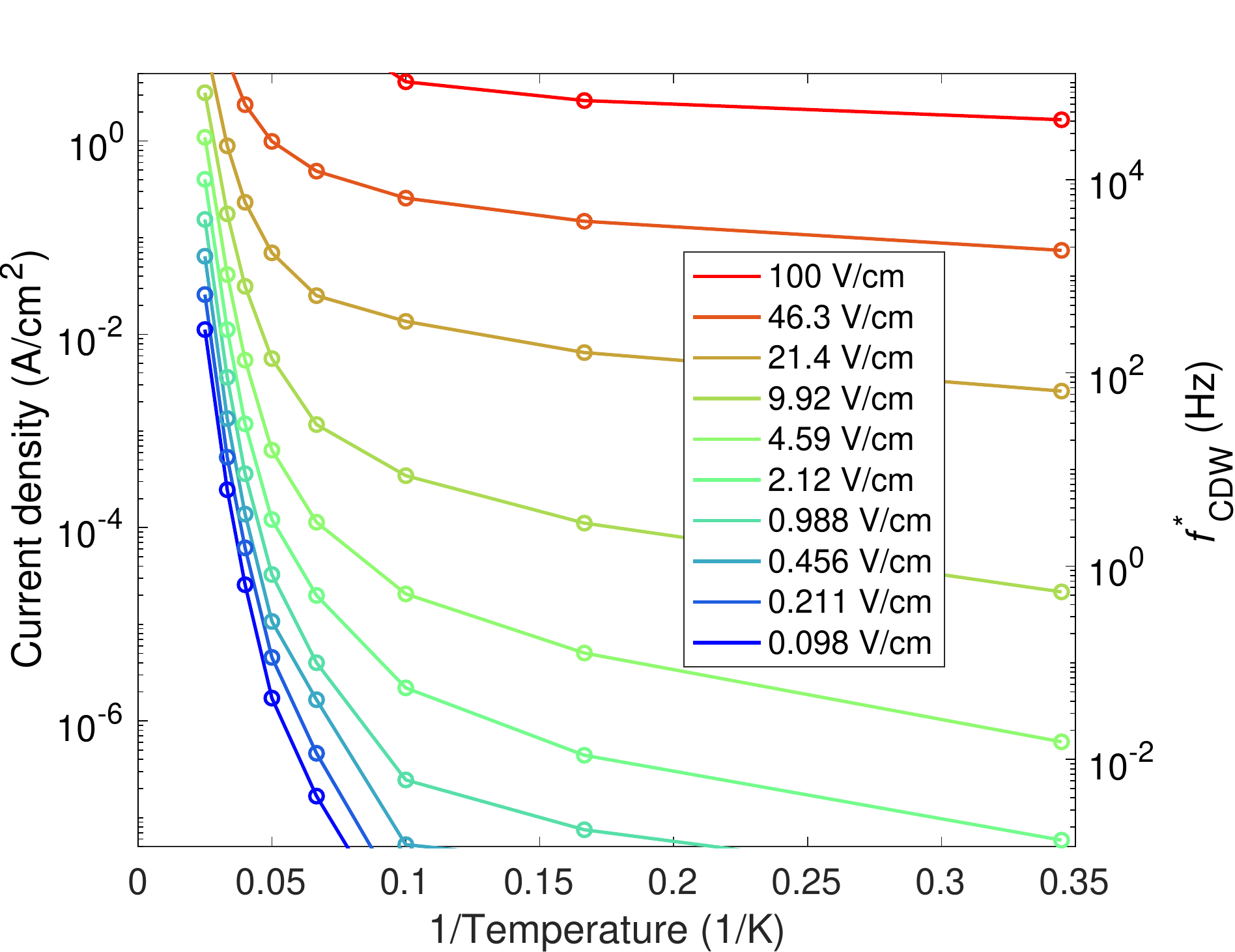}
\caption{\label{fig:vsTpure}Temperature dependences of conduction of the pure crystal in different electric fields (data of Fig.~\protect\ref{fig:ivspure}).}
\end{figure}

Respective set of temperature dependences of electric field dependent conduction is shown in Fig.~\ref{fig:vsTpure}. The transition from steep activated creep region to more flat low-temperature region is clearly seen.

In the CDW sliding mode, motion of the CDW is accompanied by generation of the narrow band noise with the frequency $f_{CDW}$ proportional to the CDW current density, $f_{CDW}=j_{CDW}/k$, where $k\approx 40$~A/cm$^2$MHz for {\it o}-TaS$_3$ \cite{GrunerReview}. In the creep mode $j\gtrapprox  j_{CDW}$, therefore motion of the CDW can be characterized by the frequency $f_{CDW}^*=j/k\gtrapprox j_{CDW}/k$ which is now the upper boundary for mean rate of overcoming of pinning barriers. The respective frequency scale is also shown in Fig.~\ref{fig:vsTpure}. In the studied samples, creep mode corresponds to $f_{CDW}^*\lesssim 100$ kHz.

\begin{figure}
\includegraphics[width=8cm]{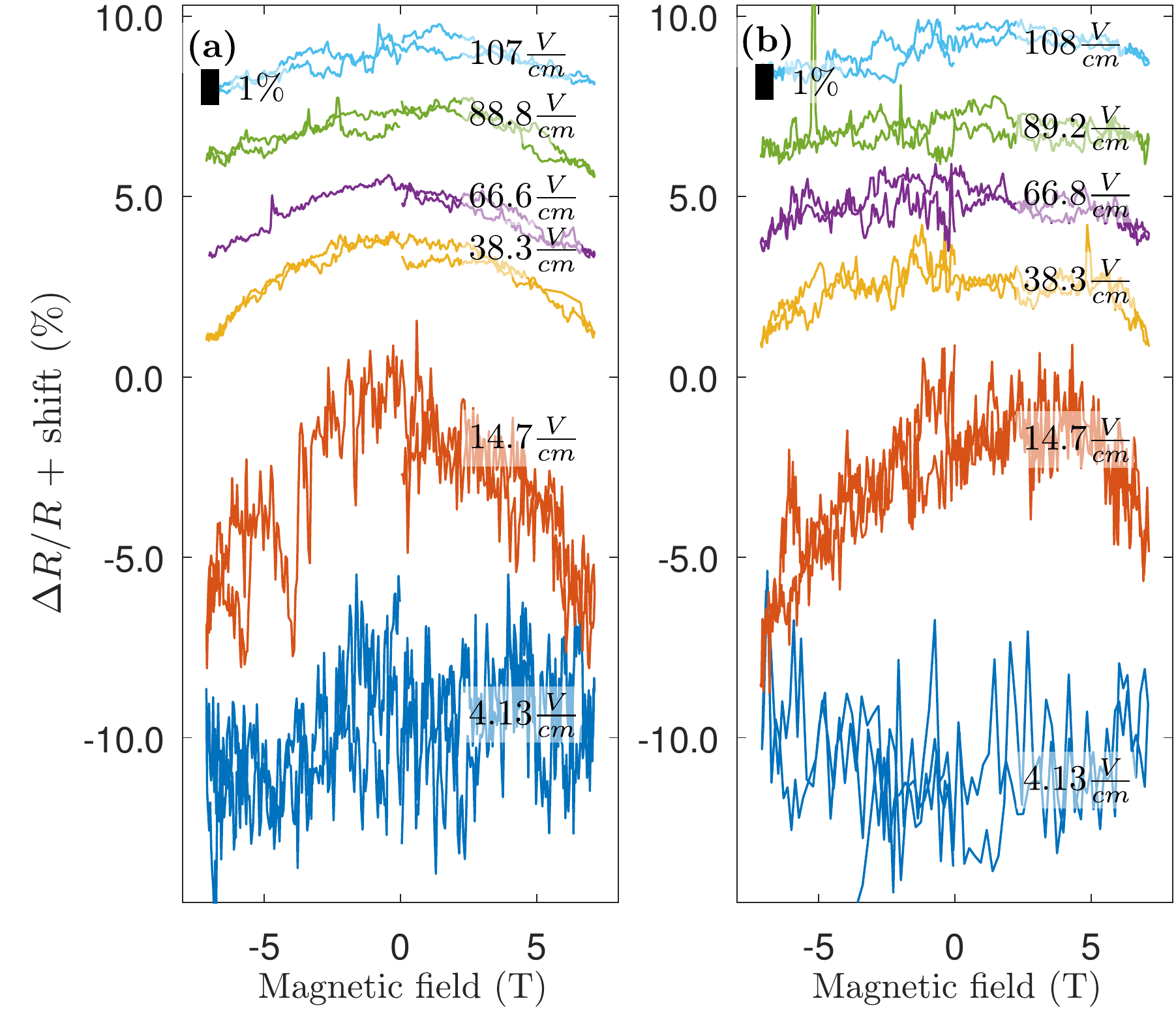}
\caption{\label{fig:bothpure}Low-temperature sets of (a) transverse and (b) longitudinal magnetoresistance in the pure sample. $T=2.9$~K.}
\end{figure}

Fig.~\ref{fig:bothpure} shows low-temperature magnetoresistance curves collected at different voltages applied to the pure sample in transverse (Fig.~\ref{fig:bothpure}(a)) and longitudinal (Fig.~\ref{fig:bothpure}(b) orientations of the magnetic field. The value of transverse negative magnetoresistance is 1.5-2 times greater the longitudinal one, but the shapes of longitudinal and transverse magnetoresistance are similar. For this sample, magnetoresistance in both orientations demonstrates the negative nearly quadratic magnetoresistance component except for the lowest electric field $E=7.29$~V/cm. The tendency for the parabolic negative magnetoresistance to transform into an almost linear one at the highest electric field is present, but not as pronounced as in less pure crystals (see below and Supplementary materials) due to broadband noise, which is especially strong in pure crystals due to the much larger value of the phase correlation length.

\begin{figure}[h]
\includegraphics[width=8cm]{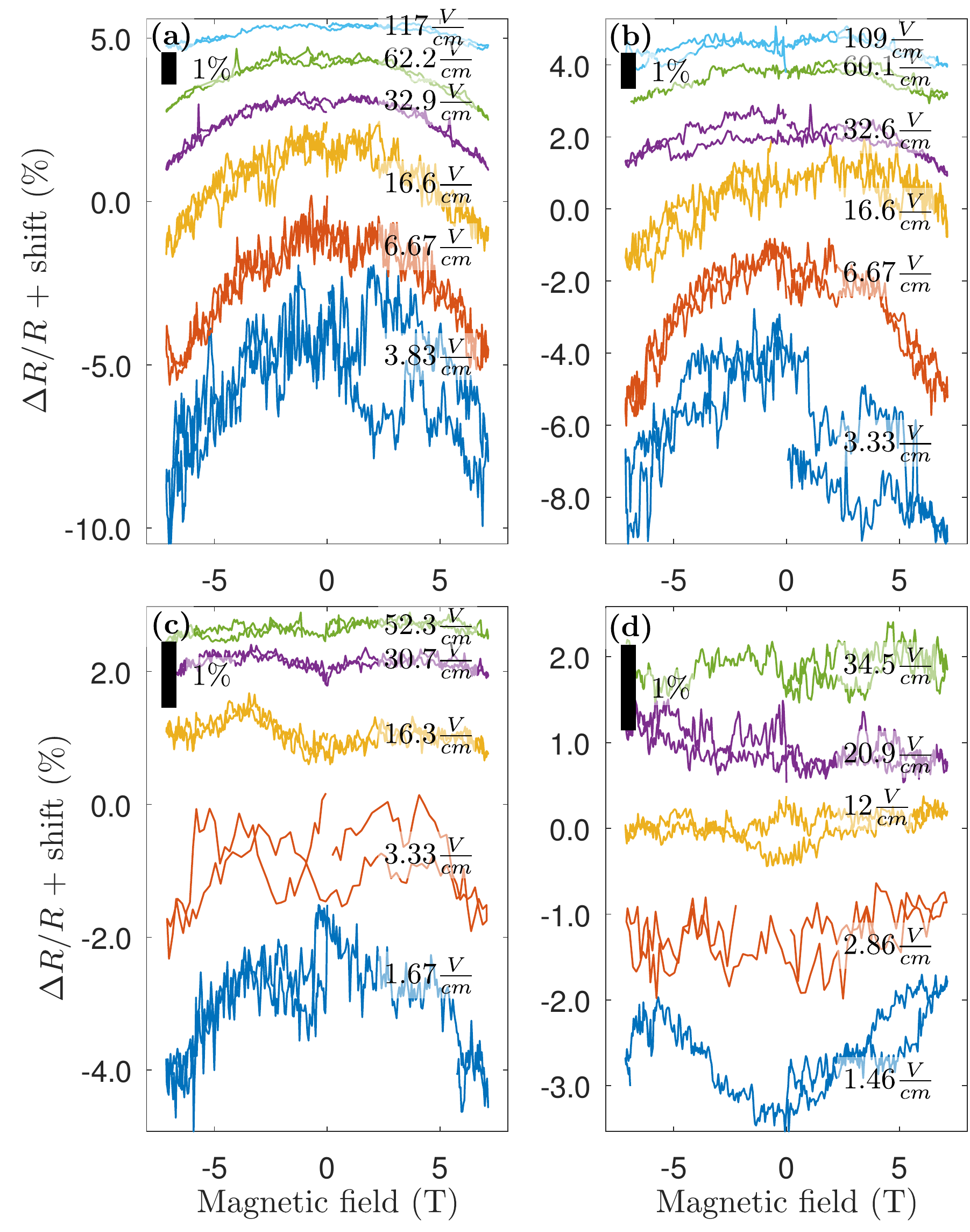}
\caption{\label{fig:allperpshort} Temperature set of electric-field dependent transverse magnetoresistance in the pure sample. a) $T=6.1$~K, b) $T=10.0$~K, c) $T=20.0$~K, d) $T = 30$~K .}
\end{figure}

 Fig.~\ref{fig:allperpshort} shows temperature sets of low-temperature transverse magnetoresistance curves collected at different voltages applied to the pure sample. Magnetoresistance is parabolic and negative at all studied temperatures and voltages applied to the sample. 
The measurement accuracy is poor due to very strong broadband noise intrinsic to pure crystals. A small  dip ($\sim 0.1$\%) of positive magnetoresistance is getting apparent with temperature increase above $T\gtrsim 20$~K and the negative magnetoresistance contribution disappears with further temperature increase. Longitudinal magnetoresistance demonstrates similar behavior but the values of magnetoresistance are 1.5-2 times smaller at all temperatures studied.  In general, in this temperature range, the observed magnetoresistance is similar to one observed in another CDW system, (TaSe$_4$)$_2$I except for the sigh of the effect which is opposite at $T>40$~K and change its sign at $T\leq 30$~K \cite{Cohn2020}. 

\subsection{Impure crystals}

\begin{figure}
\includegraphics[width=8cm]{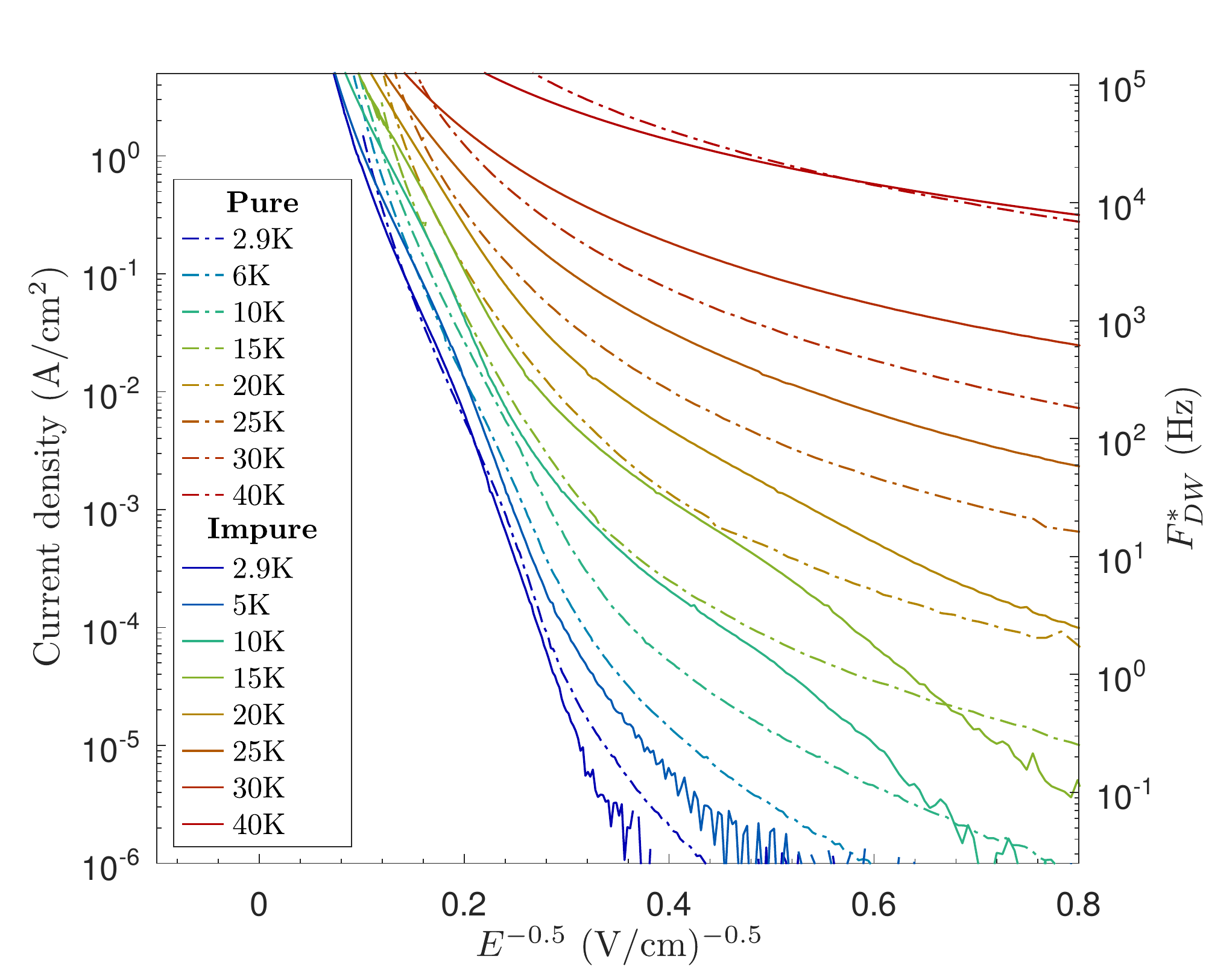}
\caption{\label{fig:ivsimpure}Low-temperature sets of {\it I-V} curves of the pure (dashed lines) and impure (solid lines) crystals in the CDW creep region.}
\end{figure}

Fig.~\ref{fig:ivsimpure} shows a low-temperature set of current-voltage characteristics ({\it I-V} curves) of both pure and impure samples plotted as $\log(j)$ vs. $1/\sqrt{E}$. The curves are typical for impure {\it o}-TaS$_3$ and consist of sections of linear conductivity (low electric field region) and CDW creep region. The nonlinear part of the curves follows $\ln{j}\propto 1/E^\mu$ law with  $\mu \approx 1/2$ law expected for thermally activated creep in 3D CDW \cite{Nattermann}, as well as for quantum creep of 1D CDW \cite{Nattermann2} (Equation \ref{eq:qqimp}). 

\begin{figure}
\includegraphics[width=8.4cm]{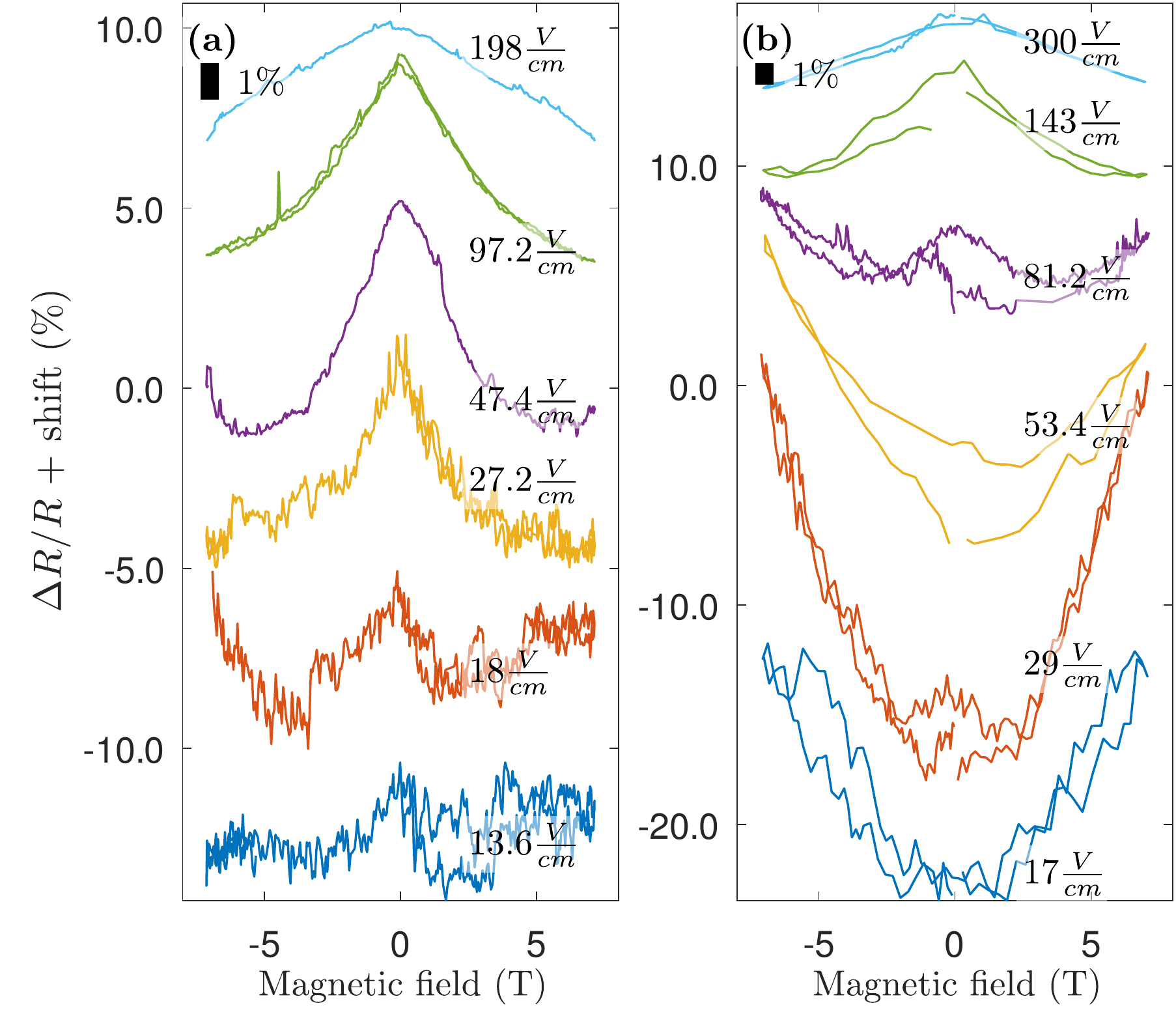}
\caption{\label{fig:longshort}Low-temperature magnetoresistance in long (a) and short (b) segments of the impure {\it o}-TaS$_3$ sample. (a) Segment length $L=5.0$~mm; (b) Segment length $L=0.21$~mm. $T = 2.9$~K.}
\end{figure}

A two-contact study leaves open the question of a possible contact contribution to the magnetoresistance. Fig.~\ref{fig:longshort} shows sets of MR curves of two segments of different length of the same crystal of {\it o}-TaS$_3$. In the long segment (Fig.~\ref{fig:longshort}(a)), positive magnetoresistance is not observed, while it prevails at the lowest electric fields in the short segment (Fig.~\ref{fig:longshort}(b)). All other features of the curves measured at close values of the current density are similar. Therefore, the main results presented below for the impure crystal are illustrated by the data obtained in the long segment.

\begin{figure}[ht]
\includegraphics[width=8.4cm]{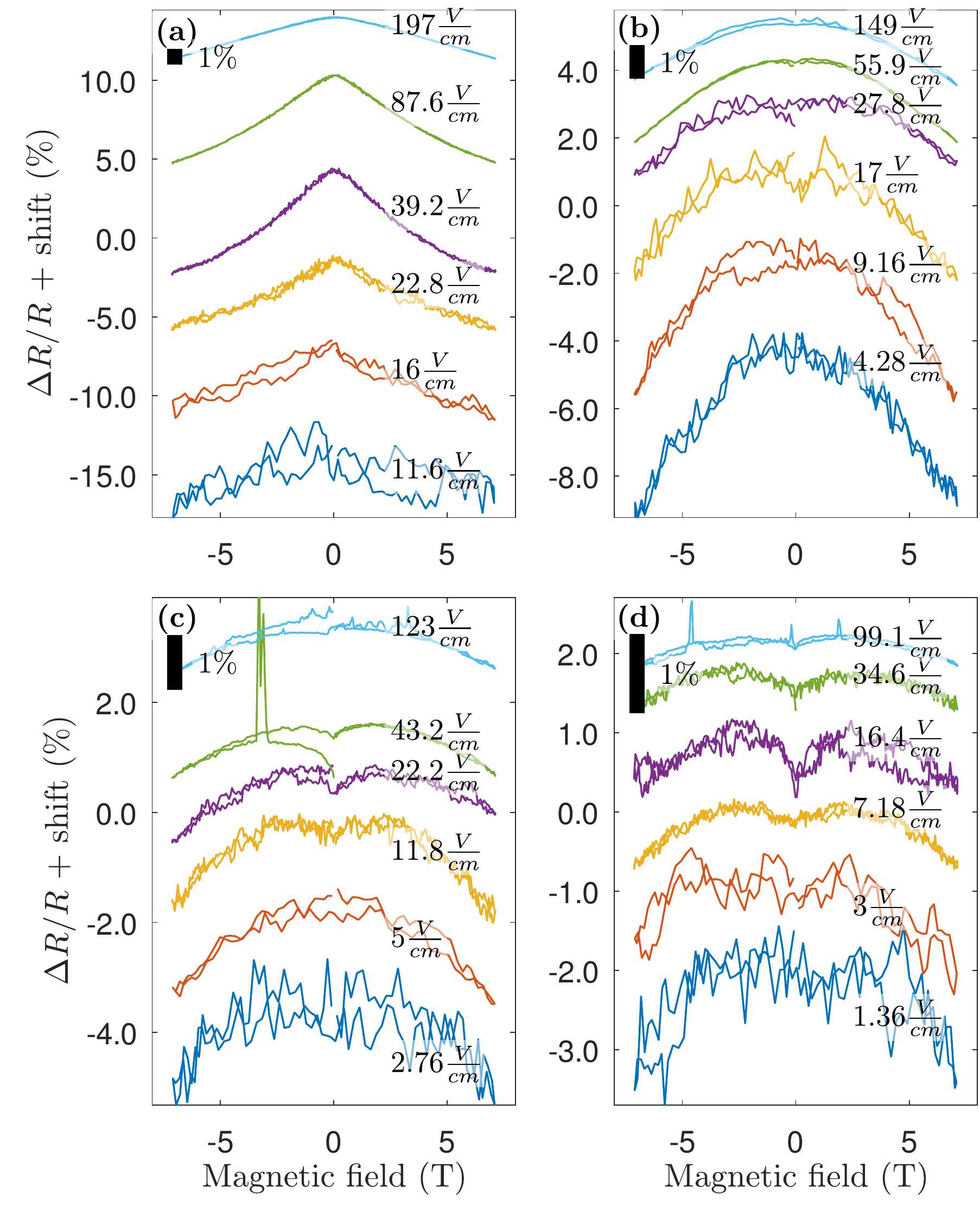}
\caption{\label{fig:perp}Temperature set of electric-field dependent transverse magnetoresistance of the impure crystal. a) $T=4.8$~K, b) $T=10.0$~K, c) $T=15.0$~K, d) $T=20.0$~K.}
\end{figure}

 Fig.~\ref{fig:perp} shows electric field sets of transverse magnetoresistance of impure sample collected at higher temperatures. The high-voltage negative nearly linear magnetoresistance  (Fig.~\ref{fig:perp}(a)) smoothly transforms into nearly parabolic one (Fig.~\ref{fig:perp}(b,c)) and then disappears with temperature increase Fig.~\ref{fig:perp}(d) above 10~K. A small ($\sim 0.1$\%) peak-like positive magnetoresistance is getting apparent with temperature increase and the negative magnetoresistance contribution disappears. The longitudinal magnetoresistance is 1.2--2 times less than the transverse one, as in the pure sample (Fig.~\ref{fig:bothpure}). Magnetoresistance in impure (Fig.~\ref{fig:perp}) and moderately pure crystals of {\it o}-TaS$_3$ (see Supplementary materials) are similar to each other qualitatively and quantitatively despite of very different transport properties. The low-temperature linear magnetoresistance is more pronounced in less pure samples.
 
\section{Discussion}
As it is noted above, 
the positive magnetoresistance observed in relatively short samples is a contact phenomena and corresponds to the presence of a dielectric barrier between the indium contacts and {\it o}-TaS$_3$. The mechanism of positive magnetoresistance for such a system is well known and corresponds to shrinking of wave functions of localized states in the dielectric barrier that results in nearly parabolic positive magnetoresistance which may be even exponentially large and has strong temperature dependence \cite{ShklEfr}. As the transparency of such a barrier is enhanced with the electric field and temperature increase, its contribution dominates at low current densities and the lowest temperatures and disappears at large ones, in agreement with our results (Fig.~\ref{fig:longshort}(b)).

\begin{figure}[h]
\includegraphics[width=8.5cm]{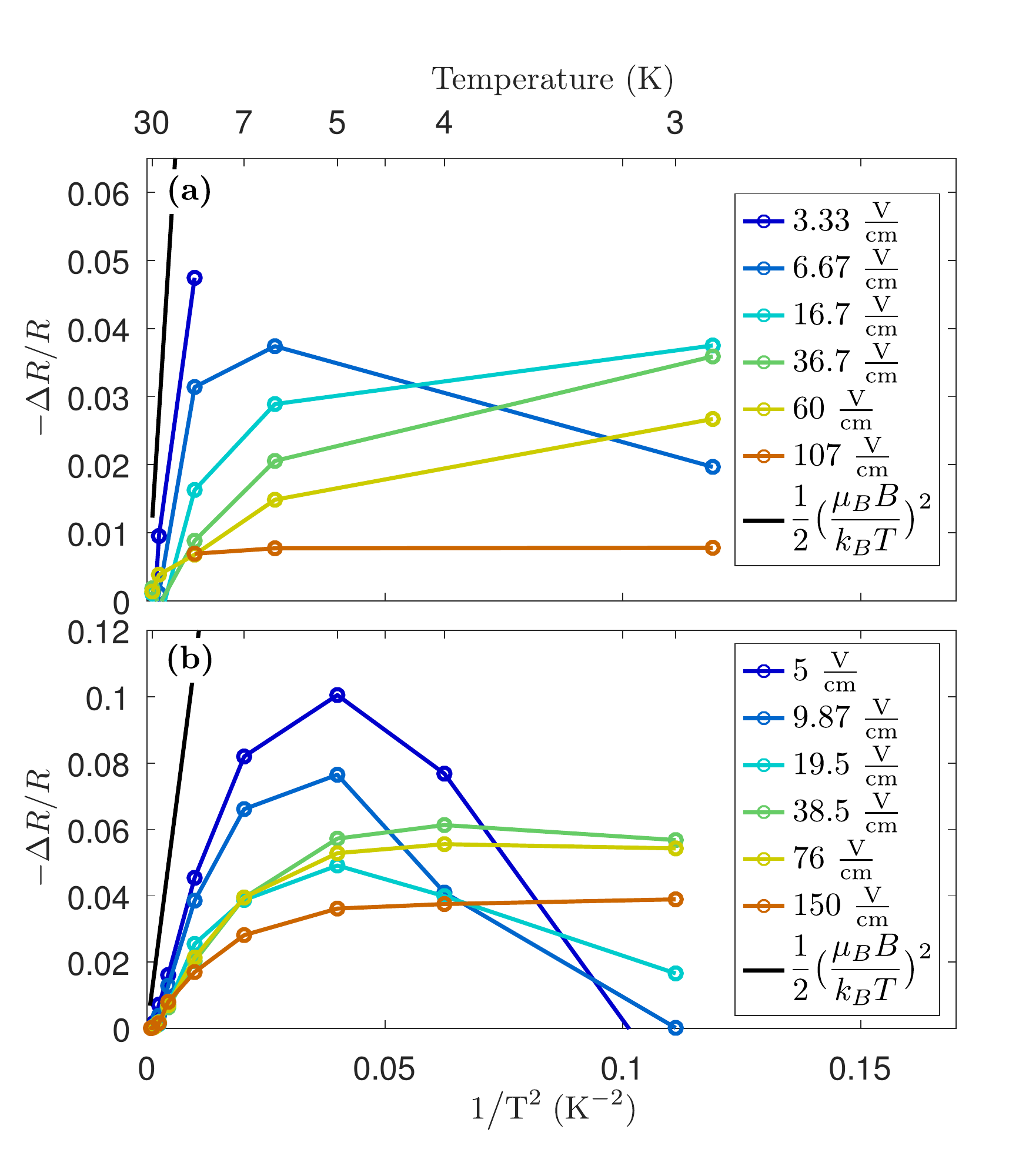}
\caption{\label{fig:dRvsTboth}The value of transverse magnetoresistance $[R(7{\rm~T})-R(0)]/R(0)$ in the pure (a) impure (b) crystals {\it vs.} $1/T^2$. Solid line shows prediction of Eq.~\ref{eq:Ohmic} with $g=1$.} 
\end{figure}

Fig.~\ref{fig:dRvsTboth}  shows 
the value of the magnetoresistance defined as $[R(7{\rm~T})-R(0)]/R(0)$ measured at various current densities as a function of $1/T^2$ in the pure (Fig.~\ref{fig:dRvsTboth}(a)) and impure (Fig.~\ref{fig:dRvsTboth}(b)) crystals. The dependences are monotonic for high electric fields $E\gtrsim 40$~V/cm and have maxima for smaller electric fields. Black solid lines in Fig.~\ref{fig:dRvsTboth} shows prediction of Eq.~\ref{eq:Ohmic} with $g=1$ which overestimates the effect by factor 3-30 (see below). We see that at least for the impure sample, the high-temperature part of the curves really follows $1/T^{2}$ dependence at $T\gtrsim 7$~K (but with electric field dependent slope) and deviates at $T\lesssim 7$~K (see also Supplementary materials). 

On the first glance, the negative quadratic magnetoresistance observed at $T\gtrsim 10$ K corresponds to Eq.~\ref{eq:Ohmic} which predicts the same scale of the effect and temperature dependences for the linear conductivity as the observed one for nonlinear one. In principle, the appearance of magnetoresistance in the nonlinear conduction could be the consequence of scaling between the linear and nonlinear conduction \cite{scaling}. We cannot accept this explanation because of the following reasons: (i) in {\it o}-TaS$_3$ the current carriers concentration at $T\sim 10$~K  is too small to provide any noticeable contribution ($\exp(E_a/T)\sim 10^{-20}$, where $E_a=450$~K is the low-temperature activation energy for linear conduction \cite{solitons}); (ii) the scaling between the linear and nonlinear conduction \cite{scaling} is not expected in the CDW creep region where the CDW current is 
determined by the rate of overcoming of pinning barriers rather than energy dissipation providing by quasiparticles or solitons and resulting in the scaling \cite{scaling}. 

\begin{figure}
\includegraphics[width=8cm]{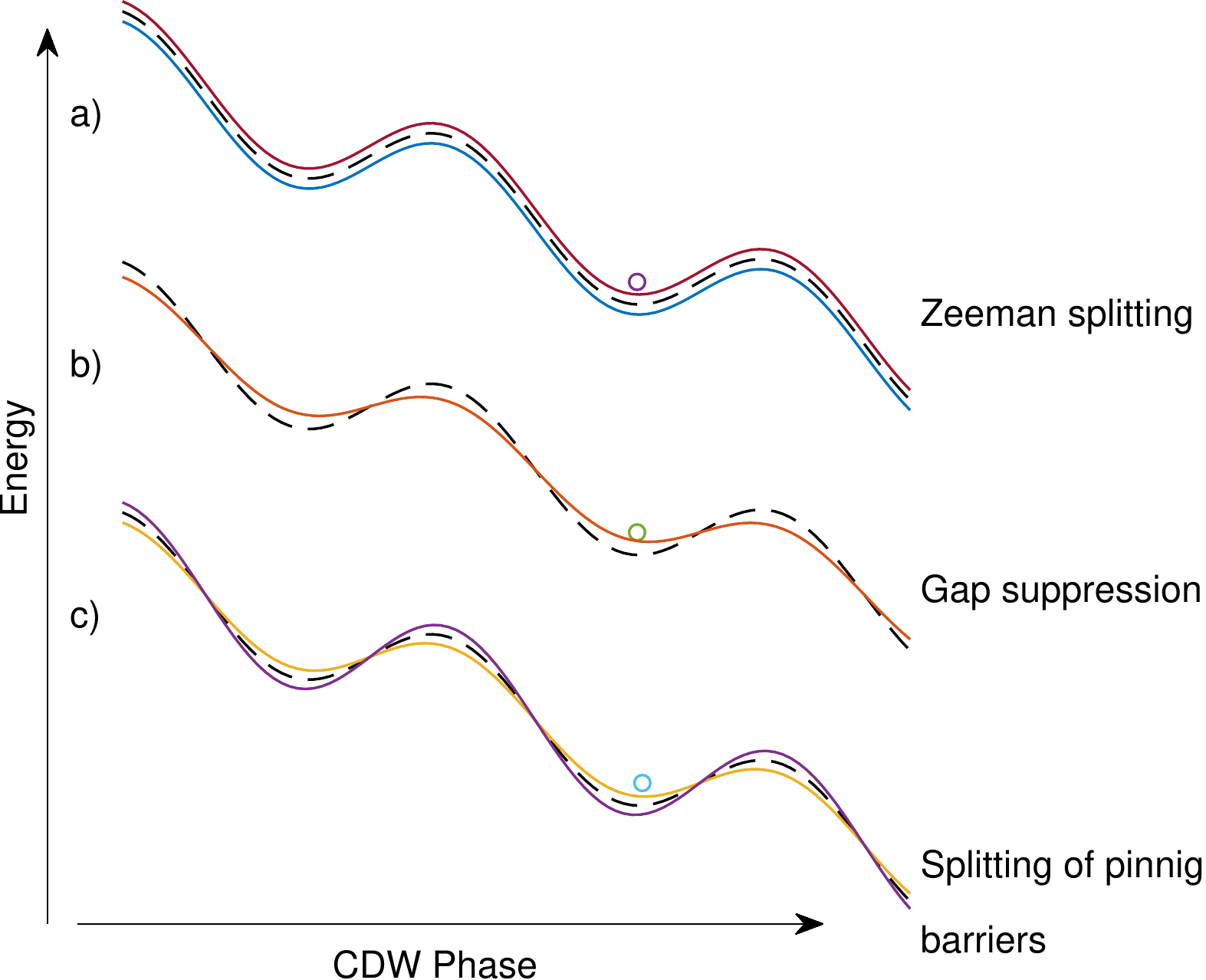}
\caption{
\label{fig:splitting}
Different possible effects of magnetic field on the pining potential in the model cosine-type of CDW interaction with impurities in the electric field $E=0.4E_T$. Dashed lines show pinning potential at $B=0$. }
\end{figure}

In the case of Zeeman splitting of pinning barriers (Fig.~\ref{fig:splitting} (a)), magnetoresistance would also follows Eq.~\ref{eq:Ohmic} which predicts $1/T^2$ temperature dependence, in agreement with the experimental results shown in Fig.~\ref{fig:dRvsTboth}. However, only the Zeeman splitting does not explain the dependence of the magnetoresistance on the electric field, which is clearly seen in Fig.~\ref{fig:dRvsTboth}.

\begin{figure}
\includegraphics[width=8cm]{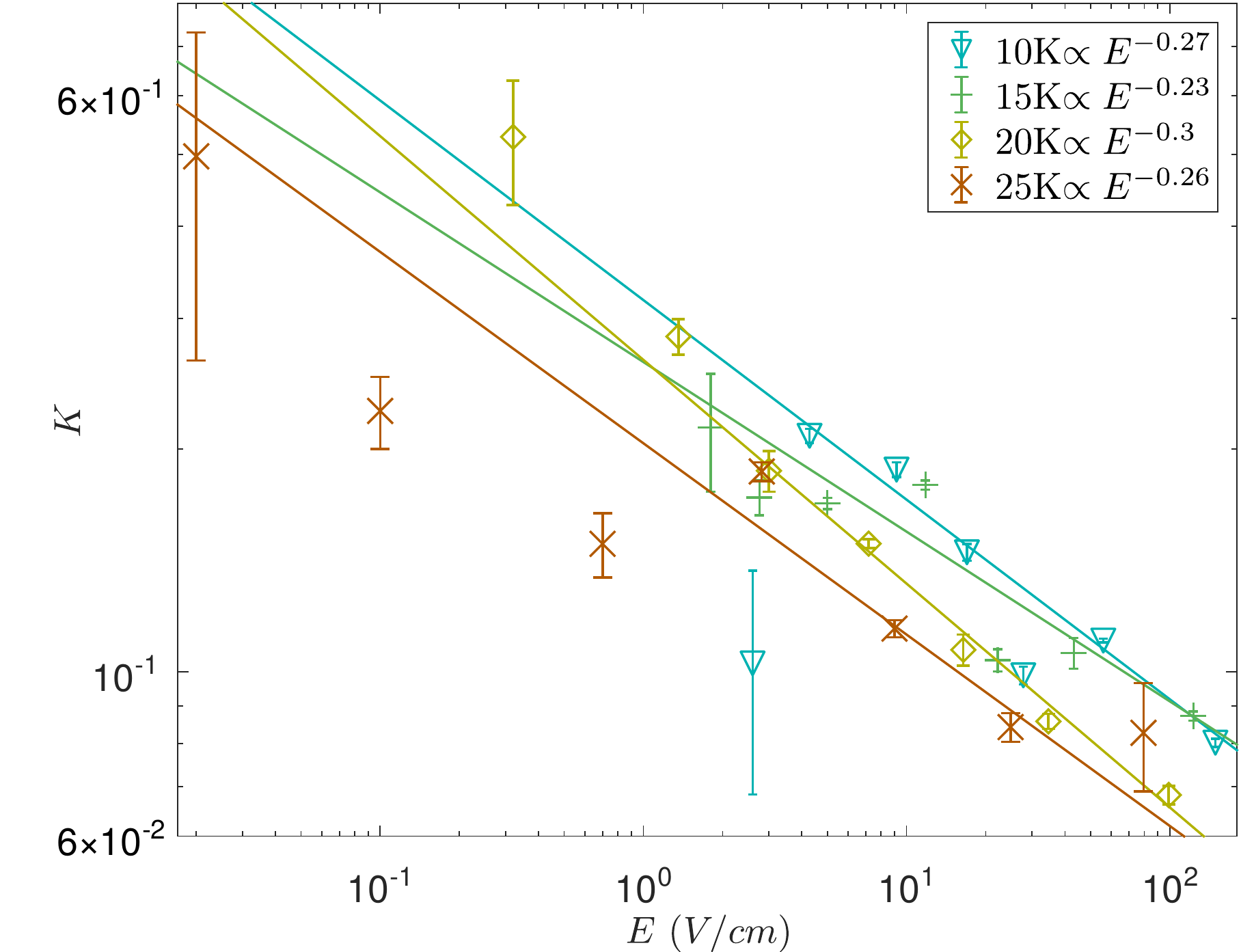}
\caption{
\label{fig:k}
Dimensionless curvature of the normalized magnetoresistance curves, $K=(k^2T^2/R(0))d^2 R(B)/(d\mu_BB)^2$, as a function of the electric field. }
\end{figure}

Fig.~\ref{fig:k} shows dimensionless curvature of magnetoresistance curves $K =-(kT^2/R(0))d^2R/d(\mu_B^2B^2)$ as a function of $E$ at different temperatures. We see that the curvature is electric field dependent and decreases with its increase. 
Since the pinning barrier heights are proportional to the CDW order parameter $\Delta$, application of magnetic field results in suppression of pinning barriers. As $T_P\propto \Delta$, the magnetic field induced suppression of the Peierls energy gap
$\delta\Delta$ follows the relation
\begin{equation}
\frac{\delta\Delta}{\Delta} = -\gamma\left(\frac{\mu_BB}{kT_P}\right)^2,
\label{eq:q}
\end{equation}
where $\gamma\sim 1$ \cite{suppression,Bjelis1999}.
 In the case of activated nonlinear conduction $R=R_0\exp{(W(E)/kT)}$, when the suppression of the order parameter (Eq. \ref{eq:q}) is the only effect of magnetic field (see Fig.~\ref{fig:splitting}(b)), then $\delta W/W =-\gamma(\mu_BB/kT_P)^2$ and the expected temperature dependence is 
\begin{equation}
\frac{\Delta R}{R}=-\gamma\frac{W(E)}{T}\left(\frac{\mu_BB}{kT_P}\right)^2,
\end{equation}
where $W(E)$ is a pinning barrier height. This equation predicts $1/T$ temperature dependence which is not consistent with our results (Fig.~\ref{fig:dRvsTboth}). Thus, the suppression itself is not enough to explain the temperature-dependent magnetoresistance.

The main contribution, consistent with our data, assumes a splitting of the pinning barriers induced by the magnetic field. In contrast to the usual Zeeman splitting, the required type of splitting depends on $E$ and diminishes when $E\rightarrow E_T$. This type of magnetic-field effect on pinning potential provides electric field dependent magnetoresistance which follows $1/T^2$ temperature dependence and follows
\begin{equation}
\frac{\Delta R}{R}=-\frac{1}{2}K\left(\frac{\mu_BB}{kT}\right)^2.
\end{equation}
The dimensionless curvature $K(E)$ is the fraction of the splitting $\mu_BB$ effective for the pinning barrier $W(E)$ in the electric field $E$. The functional form of $K(E)$ depends on the shape of pinning potential. Fig.~\ref{fig:k} demonstrates that $K$ is indeed a reasonable fraction of the usual Zeeman splitting.

The underlying physical mechanism behind the splitting of pinning barriers is not entirely clear. Generally, it corresponds to the presence of CDW subsystems with different spin order. The appearance of a low-temperature magnetic order in a CDW conductor was observed earlier in the blue bronze K$_{0.3}$MoO$_3$ \cite{BB2,BB_magnetic} though it was attributed to ferromagnetic contaminations and neglected in Ref.~\cite{BB2}. The reproducibility of the results in {\it o}-TaS$_3$ crystals of very different quality and origin makes the latter explanation unlikely. Further study is necessary to clarify the origin of this splitting. 

The transition from a high-temperature quadratic dependence to a low-temperature, close to the linear one, was observed in all four studied samples at low temperatures, regardless of their purity, and therefore is well reproducible. Note that this unusual low-temperature magnetoresistance develops {\it instead of} parabolic, rather than as an additional contribution (see Figs.~ \ref{fig:bothpure}, \ref{fig:allperpshort}, \ref{fig:longshort}, \ref{fig:perp}).

Quadratic negative magnetoresistance is observed in the temperature region of thermally activate creep. The negative linear magnetoresistance is observed in the region where the activated behavior is replaced by tunneling processes.
Unlike activated overcoming of pinning barriers, tunneling is a process that preserves the coherence of the wave function of a tunneling particle or ensemble in the case of MQT. As a result, interference of different paths must be taken into account. In the magnetic field, such interference may result in appearance of negative linear magnetoresistance, in analogy with numerous scenario for single-particle systems in the hopping conduction regime \cite{NMR_Movaghar_1978,nguen1985tunnel,hoppingNMRlin,NMR_Schirmacher}. The only difference between single-particle systems and Peierls conductors is that in the latter we are dealing with the collective motion of CDWs, and not with individual jumps of independent electrons. Quantum interference of the CDW moving  moving through columnar defects in ${\mathrm{NbSe}}_{3}$ was reported earlier \cite{Lat} (see also \cite{ZZ_review} for alternative explanation).

It is interesting to compare the results of the present study with the data obtained earlier in (TaSe$_4$)$_2$I \cite{Cohn2020}. The difference is the opposite signs of magnetoresistance between that at $T\gtrsim 20$~K in {\it o}-TaS$_3$  studied here and (TaSe$_4$)$_2$I studied earlier \cite{Cohn2020}. 
Magnetoresistance observed in both systems is consistent with the interferential effects  \cite{hoppingNMRlin,NMR_Movaghar_1978,nguen1985tunnel,NMR_Schirmacher}. The opposite signs of magnetoresistance in these system may be a consequence of strong spin-orbit interaction resulting in nontrivial topology of (TaSe$_4$)$_2$I which is expected to be a Weyl semimetal in the normal state \cite{ShiTSI}. Note that both system demonstrate inversion of the interference-like magnetoresistance  at $T^*_1\sim 15$~K ({\it o}-TaS$_3$, Fig.~\ref{fig:allperpshort}(c) and \ref{fig:perp}(c)) and $T^*_2\sim 65$~K (TaSe$_4$)$_2$I \cite{Cohn2020}, the ratio of $T^*_2/T^*_1 \approx 4$ being approximately equal to the low-temperature activation energy ratio of both systems $1800 K/450 K = 4$. Similar inversion of the sign of interference effect is known for semiconductors \cite{Studenikin}.

It should be noted that the concept of Zener CDW tunneling as a mechanism of nonlinear conduction was put forward by Bardeen \cite{Bardeen1,Bardeen2} at the early stage of the study of CDW kinetics (see Ref.~\cite{Miller} for recent results), but later gave a way to more classical phenomenological models (see Ref.~\cite{Thorne} for brief history of the concepts). The present results indicate that the time has come to take a closer look at quantum phenomena in CDW kinetics at least at low temperatures.

\section{Conclusion}

In conclusion,  we observed low-temperature magnetoresistance in {\it o}-TaS$_3$, that evolves from a  low-temperature negative linear ($T\lesssim$~10 K) to high-temperature negative parabolic one ($T\gtrsim$~10 K). The negative parabolic magnetoresistance corresponds to splitting of CDW pinning barriers by magnetic field. The linear magnetoresistance develops in the low-temperature region where quantum creep of the CDW occurs and results from CDW quantum interference similar to the scenario of negative linear magnetoresistance in single-electron systems.

{\bf Acknowledgement.} The work was supported by RScF grant \# 21-72-20114.

\bibliography{magnetores.bib}

\clearpage
\widetext

\section{Supplementary materials}
\subsection{Properties of the moderately pure sample}

\begin{figure}[h]
\includegraphics[width=8cm]{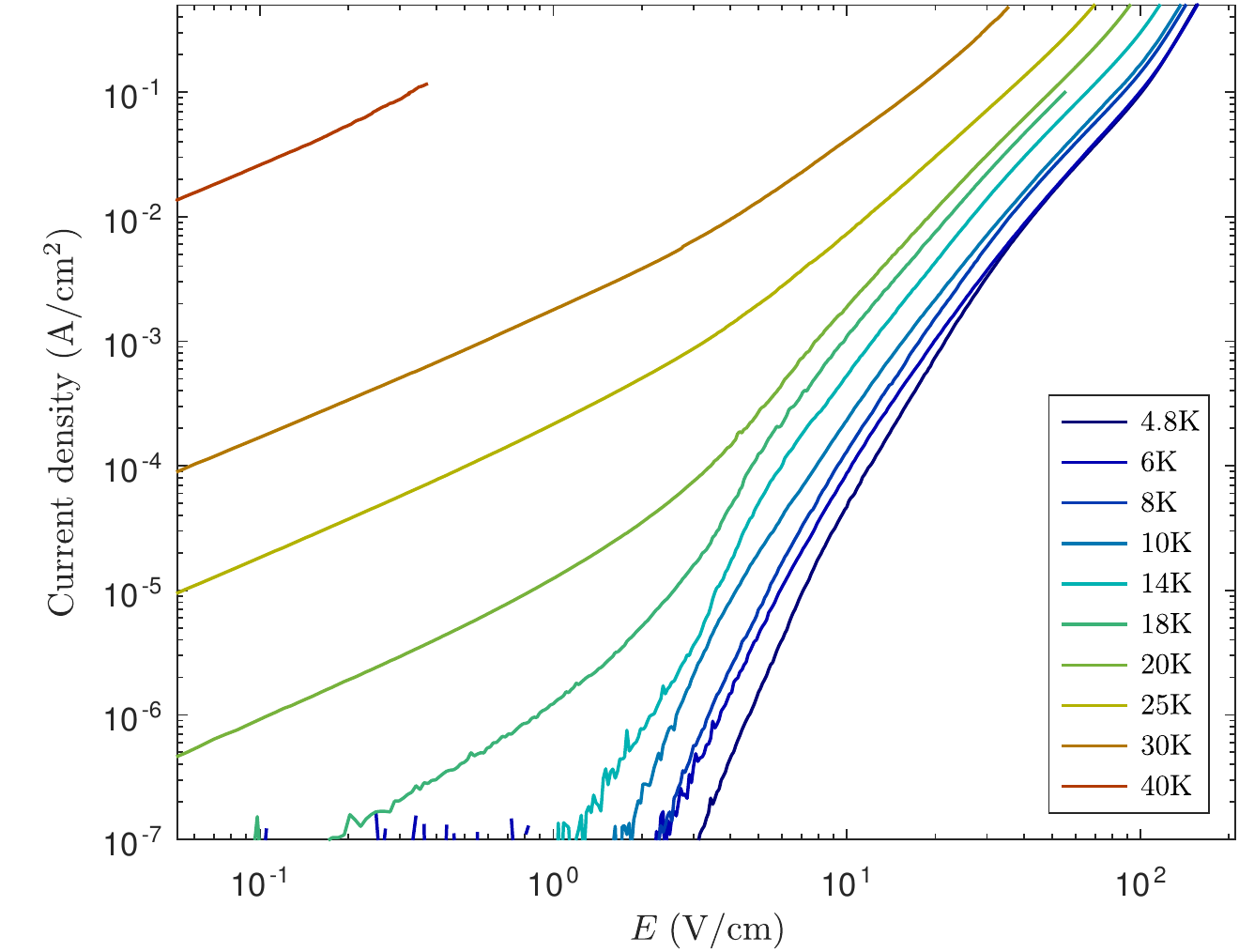} 
\caption{\label{fig:suppl_ivsmpure}Low-temperature set of {\it I-V} curves of the moderately pure crystal of  {\it o}-TaS$_3$ sample.}
\end{figure}

Here we present the data for the moderately pure crystal of o-TaS$_3$ with the following parameters: $l =1.7$~mm, $s =50\times 7.5$~$\mu$m$^2$, $V_T(100 K) = 0.6$~V. Fig.~\ref{fig:suppl_ivsmpure} shows low-temperature set of {\it I-V} of this crystal. 

\begin{figure}
\includegraphics[width=8cm]{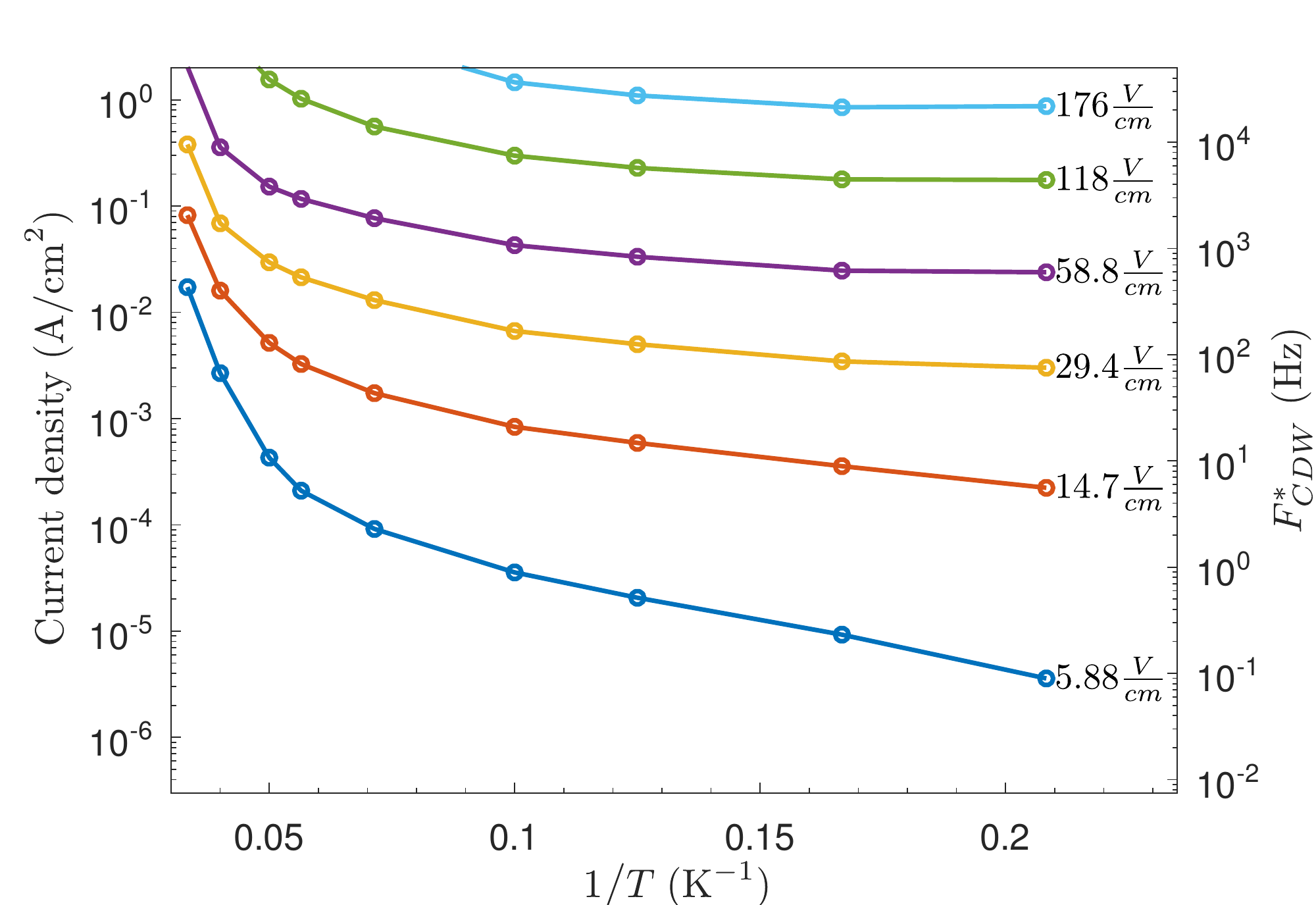}
\caption{\label{fig:suppl_vsTmpure}Temperature dependences of conduction of the moderately pure crystal in different electric fields (data of Fig.~\protect\ref{fig:suppl_ivsmpure}).}
\end{figure}

Fig.~\ref{fig:suppl_vsTmpure} shows the set of temperature dependences of conduction of the moderately pure crystal in different electric fields using the data of Fig.~\protect\ref{fig:suppl_ivsmpure}).

\begin{figure}
\includegraphics[width=8cm]{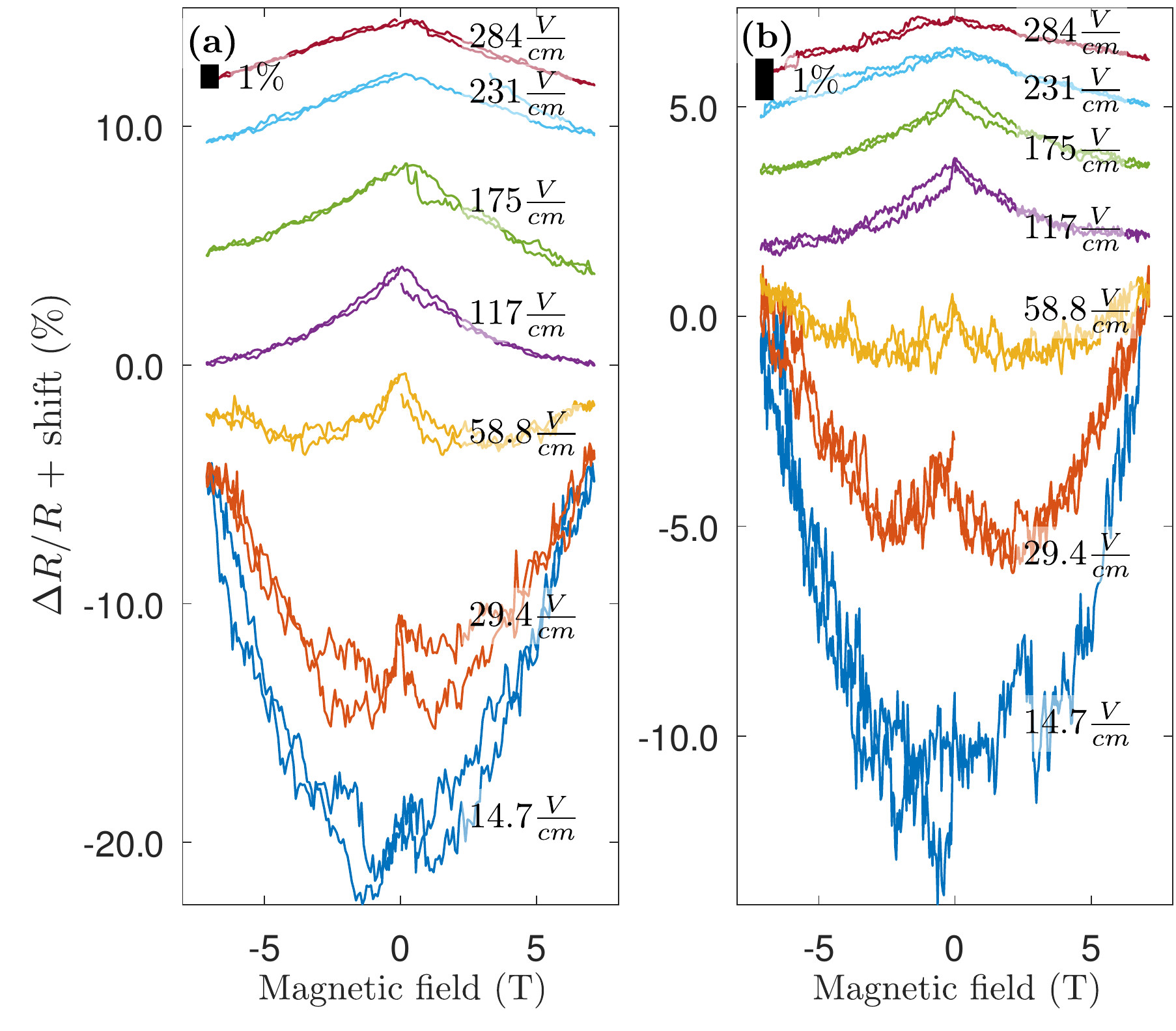}
\caption{\label{fig:suppl_bothmpure}Low-temperature sets of (a) transverse and (b) longitudinal magnetoresistance of the moderately pure sample. $T=2.9$~K.}
\end{figure}

Fig.~\ref{fig:suppl_bothmpure} shows both transverse and longitudinal magnetoresistance of moderately pure sample at the lowest temperature. 
 
\begin{figure}
\includegraphics[width=8cm]{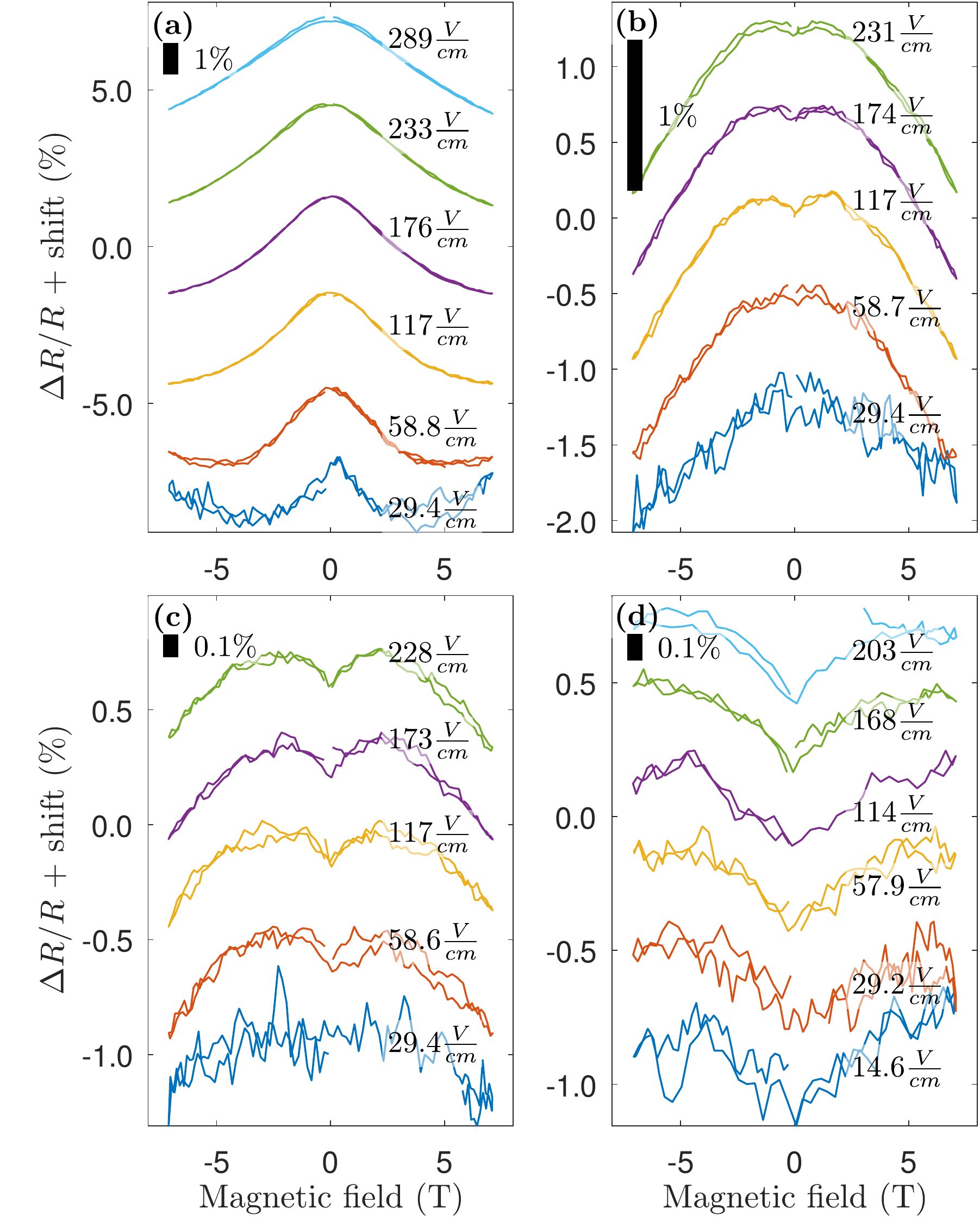}
\caption{\label{fig:suppl_mpureallperp} Temperature set of electric-field dependent transverse magnetoresistance in the moderately pure sample. a) $T=4.8$~K, b) $T=14.0$~K, c) $T=20.0$~K, d) $T = 30$~K .}
\end{figure}

\begin{figure}
\includegraphics[width=8cm]{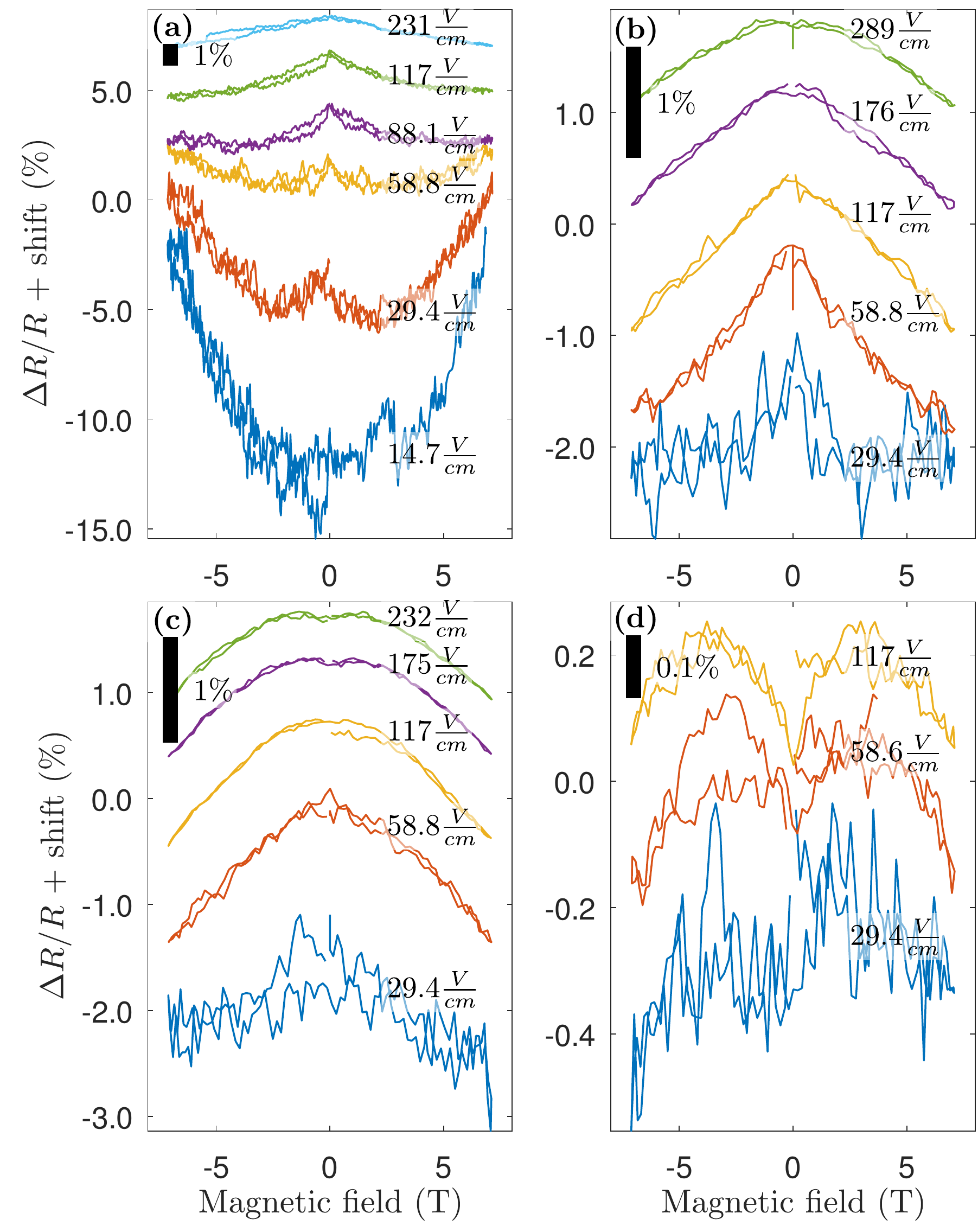}
\caption{\label{fig:suppl_mpureallpar} Temperature set of electric-field dependent longitudinal magnetoresistance in the moderately pure sample. a) $T=2.9$~K, b) $T=4.8$~K, c) $T=8.0$~K, d) $T = 20.0$~K .}
\end{figure}

Temperature sets of transverse and longitudinal magnetoresistance are given in Figs.~\ref{fig:suppl_mpureallperp} and \ref{fig:suppl_mpureallpar} respectively.

\begin{figure}
\includegraphics[width=8cm]{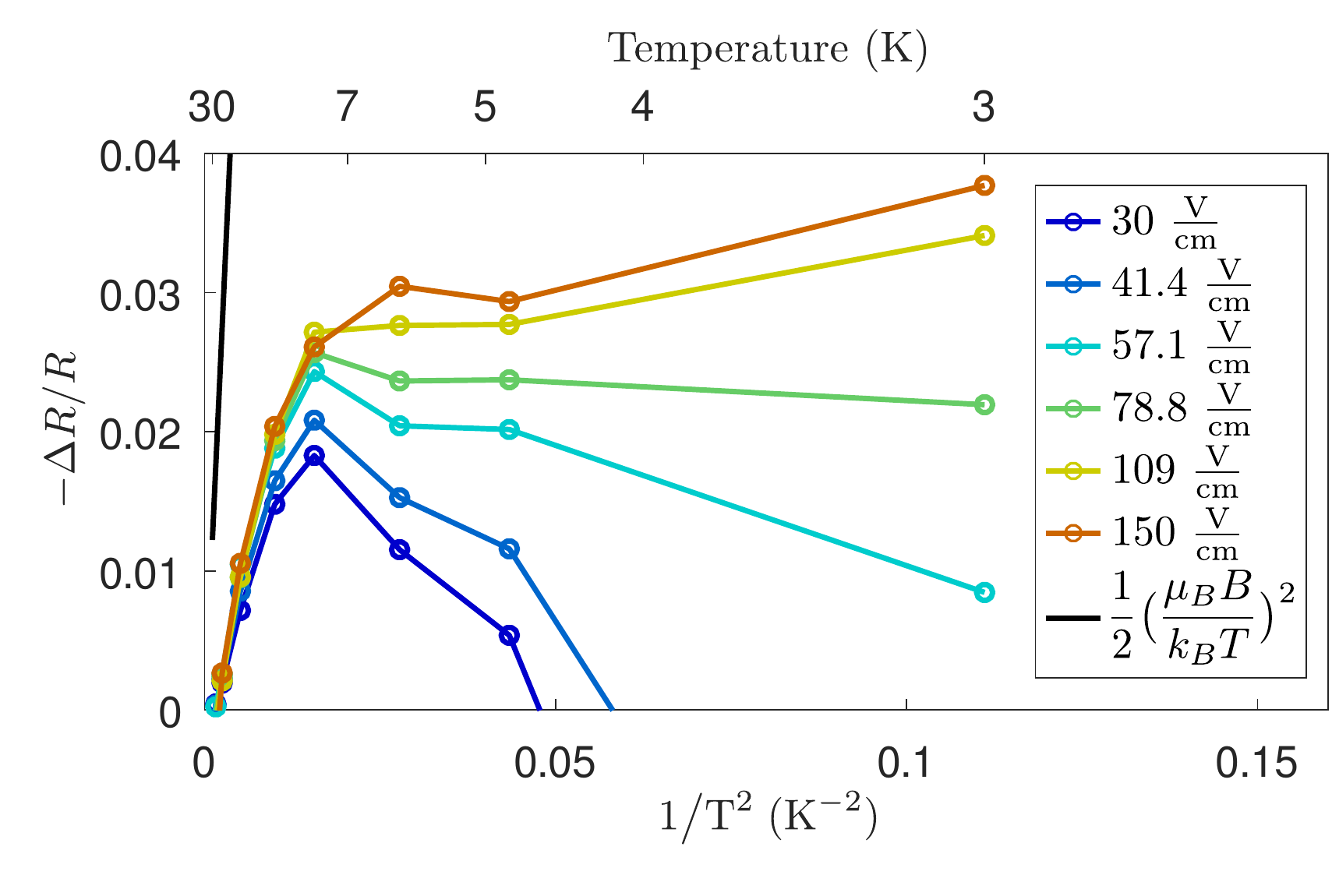}
\caption{\label{fig:suppl_dRvsT}The value of transverse magnetoresistance $[R(7{\rm~T})-R(0)]/R(0)$ in the moderately pure crystal {\it vs.} $1/T^2$. Solid line shows prediction of Eq.~\ref{eq:Ohmic} with $g=1$.} 
\end{figure}

The value of transverse magnetoresistance $[R(7{\rm~T})-R(0)]/R(0)$ in the moderately pure crystal {\it vs.} $1/T^2$ corresponding to the data shown in Fig.~\ref{fig:suppl_mpureallperp} is shown in Fig.~\ref{fig:suppl_dRvsT}.

\end{document}